\documentclass[twocolumn,superscriptaddress,showpacs,preprintnumbers,prl]{revtex4-1}

\usepackage{braket}
\usepackage{graphicx}
\usepackage{amsmath,amssymb,amstext}
\usepackage{epstopdf}
\usepackage{dcolumn}
\usepackage{hyperref}
\usepackage{dsfont}
\usepackage{gensymb}
\usepackage[dvipsnames]{xcolor}
\usepackage{nicefrac}
\usepackage[makeroom]{cancel}
\usepackage{ulem}
\usepackage{sidecap}

\newcommand*{\I}{\mathrm{i}}
\newcommand*{\E}{\mathrm{e}}

\newcommand{\modi}[2]{#2}% MODI

\begin{document}

\twocolumngrid

\title{Multifold paths of neutrons in the three-beam interferometer \\ detected by tiny energy-kick}

\author{Hermann Geppert-Kleinrath}
%\email[]{hgeppert@ati.ac.at}
\affiliation{ Atominstitut, Vienna University of Technology,
Stadionallee 2, 1020 Vienna, Austria}
\author{Tobias Denkmayr}
\affiliation{ Atominstitut, Vienna University of Technology,
Stadionallee 2, 1020 Vienna, Austria}
\author{Stephan Sponar}
\affiliation{ Atominstitut, Vienna University of Technology,
Stadionallee 2, 1020 Vienna, Austria}
\author{Hartmut Lemmel}
\affiliation{ Atominstitut, Vienna University of Technology,
Stadionallee 2, 1020 Vienna, Austria}
\affiliation{Institut Laue Langevin, 38000 Grenoble, France}
\author{Tobias Jenke}
\affiliation{Institut Laue Langevin, 38000 Grenoble, France}
\author{Yuji Hasegawa}
\email[]{hasegawa@ati.ac.at}
\affiliation{ Atominstitut, Vienna University of Technology,
Stadionallee 2, 1020 Vienna, Austria}
\affiliation{Division of Applied Physics, Hokkaido University, Kita-ku, Sapporo 060-8628, Japan}

\date{\today}

\begin{abstract}
 A neutron optical experiment is presented to investigate the paths taken by neutrons in a three-beam interferometer.
 In various beam-paths of the interferometer, the energy of the neutrons is partially shifted so that the faint traces are left along the beam-path. By ascertaining an operational meaning to "the particles's path",  which-path information is extracted from these faint traces with minimal-perturbations.
 Theory is derived by simply following the time evolution of the wave function of the neutrons, \modi{}{which clarifies the observation in the framework of standard quantum mechanics}.
 Which-way information is derived from the intensity, sinusoidally oscillating in time at different frequencies, which is considered to result from the interfering cross terms between stationary main component and the energy-shifted which-way signals. Final results give experimental evidence that the (partial) wave functions of the neutrons in each beam path are superimposed and present in multiple locations in the interferometer.
\end{abstract}


\maketitle

\section{I. INTRODUCTION}

In single-particle interference experiments, e.g., electrons, neutrons, atoms, molecules in a Mach-Zehnder type \modi{interferometers}{interferometer}, quantum interference emerges \cite{2,3,4,5,6,66}; quantum superposition, one of the most fundamental and significant features of quantum mechanics, is ascribed to this result.
The non-local coexistence of a single-particle is a consequence of simultaneous presence of a wave function in separate positions.
The capability of possessing a (non-local) wave property by massive particles \modi{was}{is} first postulated
by de Broglie in 1924 \cite{7}; the wave-particle duality is introduced and confirmed later in a diffraction experiment with electrons \cite{8}.
Feynman, in discussing the double-slit experiment, states that the single-particle interference experiments, including the wave-particle duality, has in it the heart of quantum mechanics and, in reality, it contains the only mystery \cite{1}. Optical tests of the wave-particle duality \modi{,}{and} quantum complementarity, are reported \cite{9}, which are followed by the derivation of an inequality to quantify the duality \cite{10}:
a compatible extent of (partial) fringe visibility and (partial) which-way knowledge in a two-way interferometer experiment is argued. Further on, a which-way experiment with an atom interferometer is reported \cite{11} and the validity of the inequality is confirmed in a photonic experiment \cite{12}.
\modi{}{Besides, so-called quantum-eraser experiments, where re-emergence of the interference effects by (quantum) erasing the which-way information, are reported \cite{122}.}

Recently, studies, ascertaining the photonic trajectories in a double-slit situation \cite{20} as well as photon's past in a nested Mach-Zehnder interferometer \cite{21,211,22}, are reported. The former utilizes a weak measurements of the photon's location \cite{23} to determine operationally a set of (average) trajectories for an ensemble of a photonic system.
The latter realizes (weak) which-way marking by vibrating mirrors in the interferometer. The faint traces are left on quantum particles, \modi{although}{whereas} a laser beam was used in the experiment. By ascertaining an operational and quantitative meaning to "the particle's path", the unambiguous which-path information is extracted from the faint traces \cite{BGE17};
the authors claim that the past of the photons is not a set of  continuous trajectories.
\modi{}{In addition, they assert that the two-state vector formulation provides a simple intuitive picture of the observation and that the standard quantum mechanics does not.} \cite{23}.
Along with another experimental study \cite{MB16}, the pros and cons of this claim \modi{were}{are} discussed intensively \cite{BGE17}; many-sided aspects of the observation in the experiment, are still being debated. \modi{The counterfactual reasonings of a quantum particles's trajectories remind us, to some extent, of paradoxical behavior of a post-selected quantum system, e.g., so-called Hardy's paradox; a description with a single-trajectory \`{a} la classical mechanics conflicts with the results shown in the experiments.}{}
In some arguments, the consequence of the destructive interference is overlooked
or underestimated \cite{BGE17}. Moreover, the above mentioned experiments \cite{22,MB16} \modi{were}{are} done by using a beam from a laser; as the authors admit, classical electromagnetic consideration can explain the results of the experiment.
\modi{}{It is to be noted here that identifying quantum particles' trajectories is strongly related to the issue of the interpretation of quantum mechanics \cite{Ref1,Ref3,Ref4}.}

\begin{center}
\begin{SCfigure*}
  	 \includegraphics[width=0.75 \textwidth]{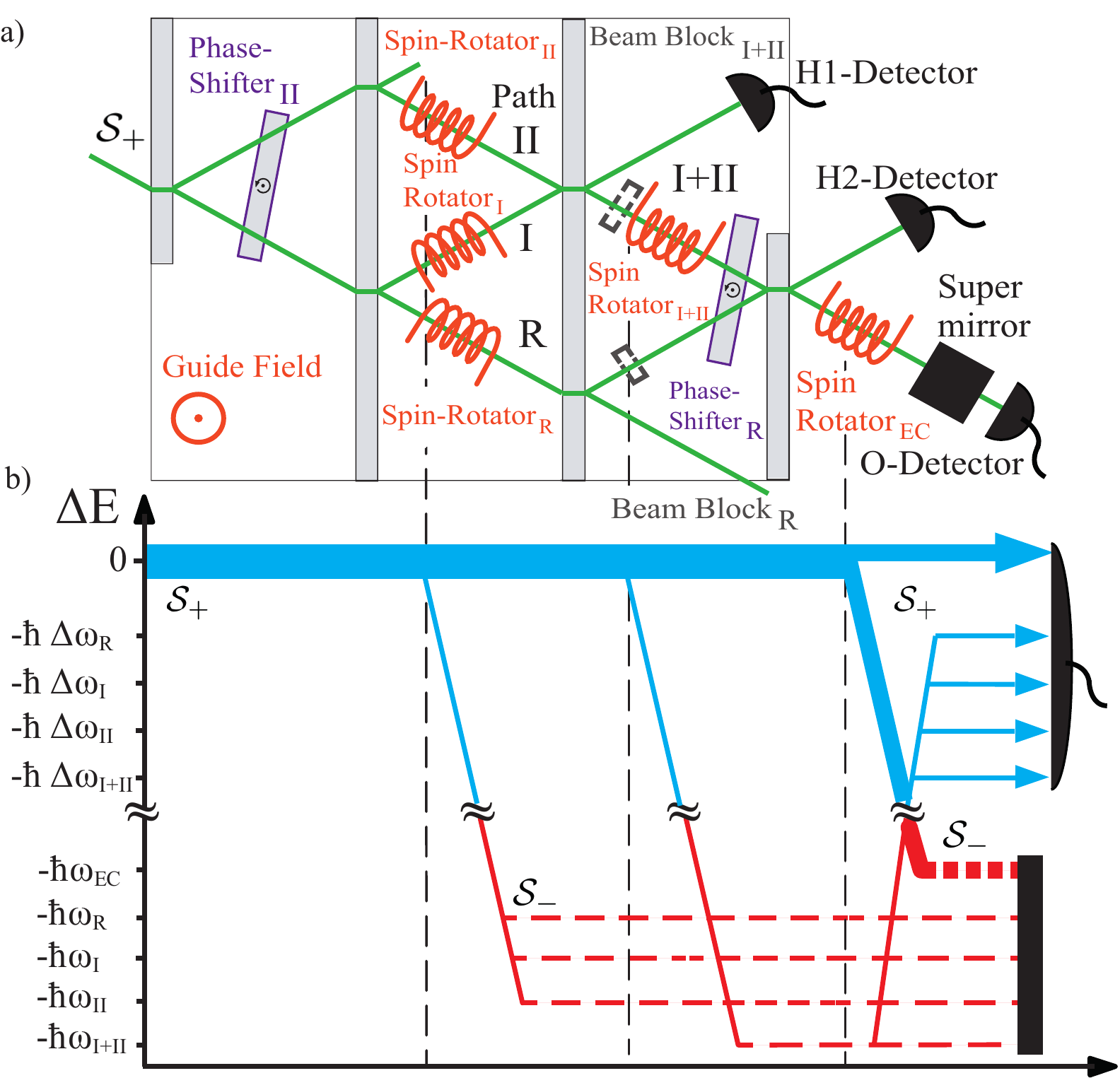}
 	 \caption{ (a) A schematic depiction of the experimental setup of the which-way measurement of neutrons in three-beam interferometer (IFM). The incident beam is split into two beam paths at the first plate of the IFM. In addition to the two interfering beams in the paths I and II,
 a reference beam is generated at the second plate of the IFM.
All three beams are recombined at the last plate of the IFM.
Between the second and the third plates, spin-rotators are inserted in each beam to accomplish which-way (WW) marking. In the recombined beam from the beams I and II, another spin-rotator perform WW-marking of the beam I+II. An energy compensation (EC) is carried out by a spin-rotator after the IFM.
After filtration of the down-spin component by a super-mirror spin-analyzer (SM), neutrons are detected by the O-detector in the forward direction.
 %Abbreviation are used such as phase shifter (PS), spin rotator (SR), guide field (GF), beam blocker (BB), energy compensation (EC), super-mirror spin-analyzer (SM) and detector (Det).%
 (b) Energy-diagram of neutrons passing through the interferometer.
 Terms higher than the first-order are neglected.
 }
\label{fig:setup}
\end{SCfigure*}
\end{center}

\section{II. THEORY}

In this \modi{letter}{paper} we present a single-neutron simultaneous (partial) which-way experiment.
Neutron interferometry \modi{has been}{is} used more than four decades to tests fundamental phenomena in quantum mechanics \cite{RW,30}; entanglement between degrees of freedom of the neutron allows investigations of quantum contextuality with massive particles \cite{Hasegawa03,Geppert14}, while the quantum Cheshire-Cat experiment is carried out recently by using a Mach-Zehnder type interferometer with neutrons \cite{Denkmayr14}.
\modi{}{In addition to the experimental demonstration of the multiple paths by the use of a \textit{pure} quantum system, instead of a laser light in a coherent state, we present here a simple theoretical treatment in the framework of standard quantum mechanics to describe the derivation of Which-way information from the faint traces.}
In the present experiment, the energy degree of freedom of neutrons is utilized to mark the neutron's paths in a three-beam interferometer (IFM) \cite{Heinrich88,Hasegawa96,Filipp05}.
The scheme of the experiment is shown in Fig.\ref{fig:setup} (a). A monochromatic neutron beam enters the IFM. The IFM consists of four plates, each working as a 50:50 beam splitter.
The IFM provides a conventional Mach-Zehnder like loop (front loop), which consists of two interfering beams in path $I$ and $II$ with the corresponding wave functions $\psi_{I}$ and $\psi_{II}$.
These two beams are recombined at the third plate of the IFM, yielding $\frac{1}{\sqrt{2}}(\E^{\I \chi_I}\psi_{I}+\E^{\I \chi_{II}}\psi_{II})=\E^{\I \chi_{I+II}}\psi_{I+II}$, where
$\chi_{I+II}=\operatorname{arg}(\E^{\I \chi_{I}}+\E^{\I \chi_{II}})$.
\modi{}{(Note that the phase factor $\E^{\I \chi_{I+II}}$ turns out to be relevant below when other beams, e.g., a reference beam, come into consideration.)}
The beam in forward direction in path $I+II$ is combined at the fourth plate of the IFM with the reference beam in path $R$, with the wave function $\psi_R$.
The wave function in path $O$ leaving the IFM in forward direction is given by
\modi{}{$\psi_O=\frac{1}{\sqrt{2}}\E^{\I \chi_{I+II}}\psi_{I+II}+\frac{1}{2}\E^{\I \chi_R}\psi_{R}$.}

The incoming neutrons are spin-polarized in positive $z$-direction represented by $s_+$, being
the eigen-state of the spin operator $\widehat{\mathcal{S}}_z=\frac{\hbar}{2}\widehat{\sigma}_z$ with the positive eingen-value, while $s_-$ denotes the eigen-state with the negative eingen-value.
The wave functions in each path in the IFM can be written in the form of spin-path wave functions with a respective phase shift as $\Psi_i= \E^{\I\chi_i} s_+ \otimes \psi_i $ ($i=I,II,R,I+II)$.
Between the second and the third plate of the IFM, the paths, taken by neutrons in the IFM, are marked by slightly shifting the energy of neutrons; the neutrons' energy serves as which-way (WW) markers \cite{12} for paths $I$, $II$, and $R$. In our experiment, WW-markings are achieved by the use of resonance-frequency spin-rotators (SR) \cite{Muskat87, Badurek83, Sponar12}.
The magnitude of the energy shift $\Delta E= \hbar \omega$ is adjusted by the frequency $\omega$ of the oscillating magnetic-field of the corresponding SR.
The amount of WW-marking is controlled by a rotation angle $\alpha$ of the neutron-spin.
This angle $\alpha$ is chosen to be small enough to minimize perturbation due to marking. The unitary transformation $\widehat{U}_{SR}(\omega, \alpha)$ represents the rotation by a SR \cite{Golub94}.
\modi{When $\widehat{U}_{SR}(\omega_i, \alpha_i)$ is applied on a wave function $\Psi_i$ the consequence is written in the form $\Psi'_i=\widehat{U}_{SR}(\omega_i, \alpha_i)  \Psi_i$ with $i=I,II,R$.}{When $\widehat{U}_{SR}(\omega, \alpha)$ is applied on a wave function $\Psi$ the consequence is written in the form $\Psi'=\widehat{U}_{SR}(\omega, \alpha)  \Psi$.}
The explicit consequences of the WW-marking of the paths I, II and R are given by,
\begin{equation}
\begin{split}
\Psi'_i&=\widehat{U}_{SR}(\omega_i, \alpha_i)  \Psi_i\\
&=\E^{\I \chi_i} \left[\cos(\alpha_i /2) s_+  - \I \E^{- \I \omega_i t}\sin(\alpha_i/2) s_-\right] \otimes \psi_i\\
&= \E^{\I \chi_i} \left[s_+  - \I \E^{- \I \omega_i t}\sin(\alpha_i/2) s_-\right] \otimes \psi_i + \mathcal{O}(\alpha^2_i),
\end{split}
\label{eq:2}
\end{equation}
with $i=I,II,R$. The energy shift results in a time dependent phase of $\E^{-\I \omega_i t}$.

The path $I+II$ coming from the front loop is marked by SR$_{I+II}$ between the third and the fourth plate of the IFM. \modi{Therefore}{Taking the amplitude reduction at the beam splitter into account}, the wave function $\Psi'_{I+II}$ behind SR$_{I+II}$ is given by
\begin{equation}
\begin{split}
\Psi'_{I+II}&=\widehat{U}_{SR}(\omega_{I+II}, \modi{}{\alpha_{I+II}})\frac{1}{\sqrt{2}}(\Psi'_I+\Psi'_{II})\\
&= s_+ \otimes \psi_{I+II}\\
& - \frac{\I}{\sqrt{2}}\sum\limits_{i} \sin(\alpha_i/2) \E^{-\I \omega_i t} \E^{\I \chi_i} s_- \otimes \psi_i+ \mathcal{O}(\alpha^2_i).
\end{split}
%\Psi'_{I+II}=\widehat{U}_{SR}(\omega_{I+II}, \alpha_{\modi{}{I+II}})\frac{1}{\sqrt{2}}(\Psi'_I+\Psi'_{II}).%
\label{eq:3}
\end{equation}
\modi{with $i=I,II,I+II$.}{} This beam is further combined with the reference beam $\Psi'_R$ at the last plate of the IFM. Thus, the wave function in path $O$ behind the IFM is given by $\Psi'_O= \frac{1}{\sqrt{2}} \Psi'_{I+II}+ \frac{1}{2}\Psi'_R$.

Behind the IFM an energy compensation of $\Delta E_{EC}=\hbar \omega_{EC}$ is performed by SR$_{EC}$ in path $O$ to reduce the overall energy shift of the WW-marking. This is achieved by a spin-rotation of $\alpha^{\pm}_{EC}=\pm\pi/2$ at a frequency of $\omega_{EC}$. The wave function behind SR$_{EC}$ is given by $\Psi^{\pm}_{EC}=\widehat{U}_{SR}(\omega_{EC},\alpha^{\pm}_{EC})\Psi'_O$. Before the beam reaches the O-detector a super-mirror spin-analyzer filters out the down-spin component, which allows to resolve sinusoidal intensity-modulations in time. The function of the super-mirror is represented by the projector $\widehat{\Pi}_{s_+}$ onto the ${s_+}$ state. For simplicity, we set the same spin-rotation angle $\alpha_i=\alpha_{ww}$ for all SR$_i$ \modi{}{($i=I, II, I+II, R$), which are used for WW-marking}; the wave function behind the super-mirror $\Psi^{\pm}_{SM}= \widehat{\Pi}_{s_+}\Psi^{\pm}_{EC}$ is then written in the form
%\begin{equation}
%\Psi^{\pm}_{SM}=\Psi_{E_0}+\sum\limits_{i} \Psi^{\pm}_{i} + \mathcal{O}(\alpha^2_{ww}),
%\end{equation}
\begin{widetext}
\begin{eqnarray}
\Psi^{\pm}_{SM}&=&  \left[ \frac{\E^{\I \chi_I}}{2\sqrt{2}}\left[ 1\mp \sin(\frac{\alpha_{ww}}{2}) (\E^{-\I \Delta\omega_I t} + \E^{-\I \Delta\omega_{I+II} t})\right] \right. \psi_I \nonumber
 +\frac{\E^{\I \chi_{II}}}{2\sqrt{2}}\left[ 1\mp \sin(\frac{\alpha_{ww}}{2}) (\E^{-\I \Delta\omega_{II} t} +\E^{-\I \Delta\omega_{I+II} t})\right]\psi_{II} \nonumber \\
&\space& +\left. \frac{\E^{\I \chi_R}}{2\sqrt{2}}\left[ 1 \mp \sin(\frac{\alpha_{ww}}{2}) (\E^{-\I \Delta\omega_{R} t} \right] \psi_{R}) \right] \otimes s_+ + \mathcal{O}(\alpha^2_{ww}) \nonumber  \\
&=&\Psi_{E_0}+\sum\limits_{i} \Psi^{\pm}_{i} + \mathcal{O}(\alpha^2_{ww}),
\end{eqnarray}
\end{widetext}
with the energy-unshifted main component
\modi{$\Psi_{E_0}=\sum\limits_{i}  \frac{\E^{\I \chi_i}} {2\sqrt{2}} \Psi_{i}$,}
     {$\Psi_{E_0}=\sum\limits_{i}  \Psi_{i}$} $(i=I,II,R)$ and
the energy-shifted components $\Psi^{\pm}_{i}$, i.e. the WW-signal; these are given  by
\begin{equation}
\begin{split}
\Psi^{\pm}_{i}=\mp \frac{1}{2\sqrt{2}} \E^{\I \chi_i} \sin\left(\frac{\alpha_{ww}}{2}\right)
\E^{-\I \Delta\omega_i t} s_+\otimes\psi_i,
\end{split}
\label{eq:6}
\end{equation}
for ${i=I,II,R,I+II}$.
The intensity at the O-detector can be calculated, up to the first order of $\alpha_{ww}$, by summing up the (stationary) intensity from the energy-unshifted main component $\Psi_{E_0}$ and that oscillating in time from the individual cross terms between the main component $\Psi_{E_0}$ and the marking components $\Psi^{\pm}_i$, \modi{as given in}{which is evaluated as}
\begin{equation}
\begin{split}
I^{\pm}=|&\Psi^{\pm}_{SM}|^2 \\
=|&\Psi_{E_0}|^2 + 2\sum\limits_{i} \mathfrak{Re} \left( \Psi^{\ast}_{E_0}  \Psi^{\pm}_{i} \right)+\mathcal{O}(\alpha^2_{ww}),
\end{split}
\label{eq:7}
\end{equation}
for $i=I,II,R,I+II$.

In our experiment, particular attention is paid on two phase settings, (i) $\chi_{II}=0$ and (ii) $\chi_{II}=\pi$, both with $\chi_{I}=\chi_R=0$. The intensities at the O-detector are calculated as
\begin{eqnarray}
I^{\pm}&(\chi_{II}&=0,\chi_R=0)=\frac{1}{32} \bigg[9 \mp 6 \sin\left(\frac{\alpha_{ww}}{2}\right) \big[  \cos(\Delta\omega_{I} t) \nonumber \\
&+&\cos(\Delta\omega_{II} t)+\cos(\Delta\omega_{R} t)+2  \cos(\Delta\omega_{I+II} t)\big]\bigg],
\label{eq:8} \\
I^{\pm}&(\chi_{II}&=\pi,\chi_R=0)=\frac{1}{32} \bigg[1 \mp 2 \sin\left(\frac{\alpha_{ww}}{2}\right) \big[  \cos(\Delta\omega_{I} t) \nonumber \\
&-&\cos(\Delta\omega_{II} t)+\cos(\Delta\omega_{R} t)\big]\bigg].
\label{eq:9}
\end{eqnarray}
Note that the (stationary) mean intensity \modi{$I^{\pm}_0=|\Psi_{E_0}|^2$}{$I^{\pm}_{E_0}=|\Psi_{E_0}|^2 =
|\sum\limits_{i} \Psi_{i}|^2$} \modi{gets}{is reduced to} $1/9$ by changing  $\chi_{II}=0$ to $\pi$ \modi{, since}{;} the amplitude of $\Psi_{E_0}$ becomes $1/3$ as  $\Psi_I+\E^{\I \pi}\Psi_{II}+\Psi_R=\frac{1}{3}(\Psi_I+\Psi_{II}+\Psi_R)$.

Since each path is marked with a different energy shift $\Delta E_i= \hbar \omega_i$, WW-information can be derived by a Fourier-analysis of the time spectrum obtained at the O-detector. If a Fourier component corresponding to a frequency $\Delta \omega_i$ is found, \modi{it is clear evidence that the neutrons have}{this is clear evidence of neutrons having} interacted with the respective $SR_i$. For instance, Eqs. \ref{eq:8} and \ref{eq:9}, suggest that neutrons have taken the paths $I$, $II$, and $R$ for both settings but that the path I+II has not been taken for the latter.

An energy-diagram of neutrons passing through the three-beam interferometer with which-way markings
is shown in Fig.\ref{fig:setup} (b). Here,  terms higher than the first-order are neglected. Up-spin components $s_+$ are represented by the solid blue line, and down-spin components  $s_-$ by the dashed red line. Only the up-spin components reach the O-detector; the down-spin components are filtered out. It should be emphasized here that
the interfering cross terms between stationary main component (blue thick line) and the energy-shifted which-way signals (blue thin lines) are responsible for the sinusoidal intensity-modulation in time
at the O-detector; which-way information is extracted from these oscillating intenisties.

\section{III. EXPERIMENT}
\subsection{A. Experimental procedure}

The experiment was carried out at the S18 instrument at  the research reactor at the Institute Laue-Langevin (ILL) in Grenoble, France. A monochromator selects neutrons with a mean wave length of
$\lambda=1.92(2)$\, \AA; the spin is polarized by two birefringent magnetic prisms.
To avoid depolarization a magnetic guide field of $25$\,G is applied over the whole setup.
Since only relative phases between corresponding beams are relevant,
only two phase shifters PS$_{II}$ and PS$_R$ are used.
The phase of path $I$ is set to $\chi_I=0$. The phase shifter PS$_{II}$ tunes the phase $\chi_{II}$ of path $II$ relative to path $I$ and therefore controls the front loop, which is monitored by the H1-detector. The phase shifter PS$_R$ tunes the phase $\chi_R$ of the reference beam, which is monitored using the H2-detector. In our experiment, the phase shifter PS$_R$ is set to give $\chi_R=0$ for all measurements. All three detectors are $^3$He counter tubes with a very high efficiency \modi{}{($>99\% $) \cite{RW,30}}. A beam-blocker can be inserted into
the IFM at two positions ${Beam Block}_{I+II}$ $(BB_{I+II})$ and ${Beam Block}_{R}$ $(BB_R)$, in beam path $I+II$ and path $R$ respectively.

The paths, taken by neutrons, are marked at different frequencies $\omega_{I}=74$\,kHz, $\omega_{II}=77$\,kHz, $\omega_{I+II}=80$\,kHz, and $\omega_{R}=71$\,kHz respectively.
All rotation angles for the WW-marking SRs are set to $\alpha_{ww}=\pi/9$.
Note that the wave functions before and after these SRs are still overlapping by $0.98$, which justifies the condition of minimal perturbation due to WW-marking.
The energy compensating $SR_{EC}$ is set to the frequency $\omega_{EC}=68$\,kHz.
When the intensity difference $\Delta I=I^+-I^-$ is calculated, the stationary part cancel out
and the element oscillating in time remains.

\subsection{B. Time spectrum}
\label{sec:TS}

\begin{figure}
 	 \includegraphics[width=0.5 \textwidth]{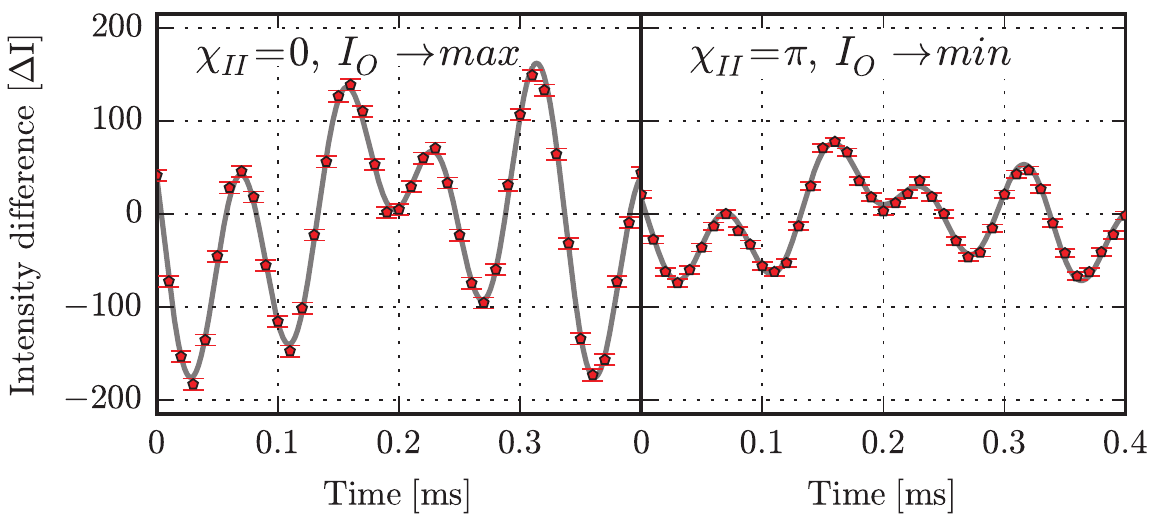}
 	 \caption{(color online).  Time spectra of measured intensity difference \modi{$\Delta I$}{$\Delta I = I^+-I^-$} for phase shifter position$\chi_{II}=0$ (left) and $\pi$ (right), together with least-squares fits.}
\label{fig:sine}
\end{figure}

In the measurements of the time spectrum,
neutrons collected for half of the time give the intensity $I^+$ and the other half of the time $I^-$.
We set the measurement time of the time spectrum for one set of phase settings
typically for $24$ to $50$\,hours.
In Fig.\ref{fig:sine} time spectra of the measured intensity differences for phase shifter position $\chi_{II}=0$ (on the left) and $\pi$ (on the right) are depicted,
together with least-square fits of four overlayed sine waves.

\subsection{C. Results}

\begin{figure}
 	 \includegraphics[width=0.5 \textwidth]{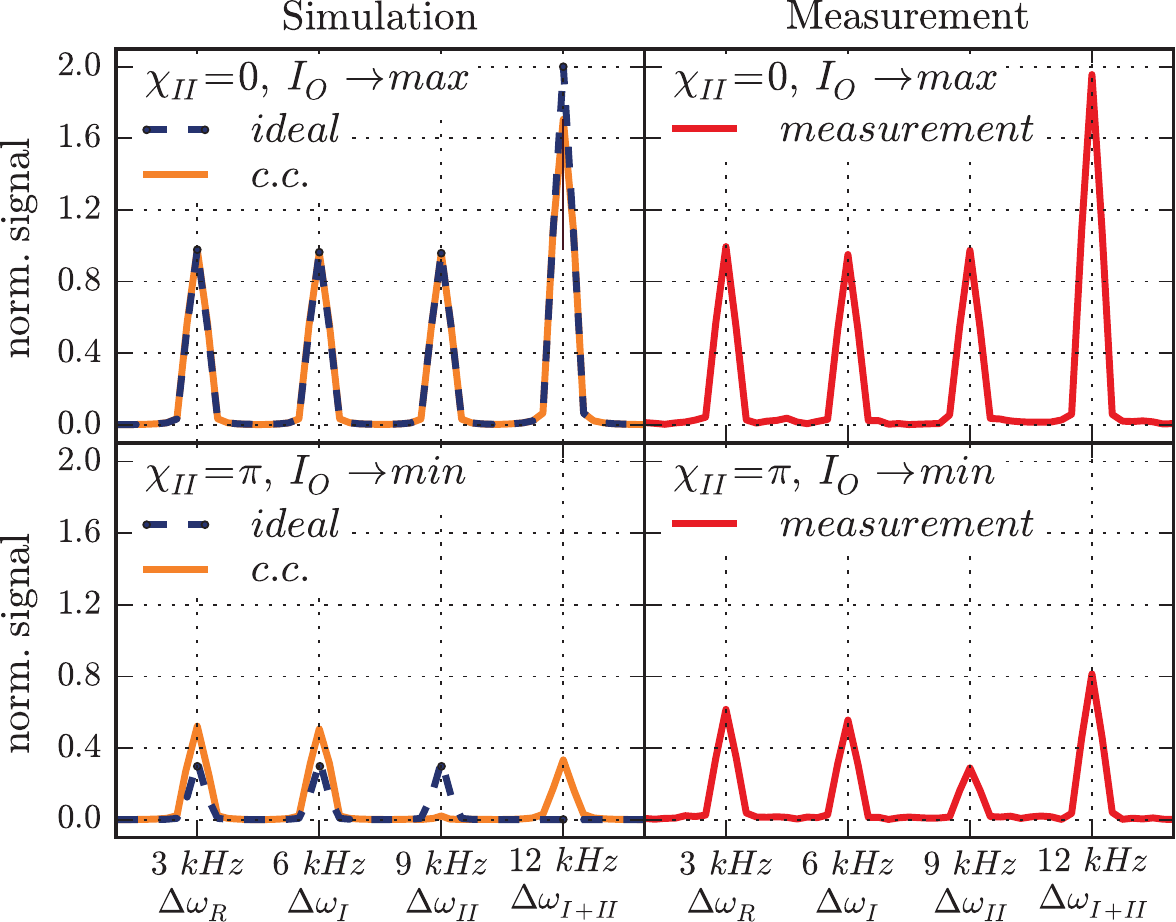}
 	 \caption{(color online). Power spectra of simulations for perfect contrast (dashed blue line) and contrast corrected (solid orange line) are plotted on the left. Power spectra of the measurement (solid red line) are plotted on the right.
 The first row is obtained for the phase setting $\chi_{II}=0$, while the second row for the setting $\chi_{II}=\pi$.}
\label{fig:NullPi}
\end{figure}

\begin{figure}
 	 \includegraphics[width=0.5 \textwidth]{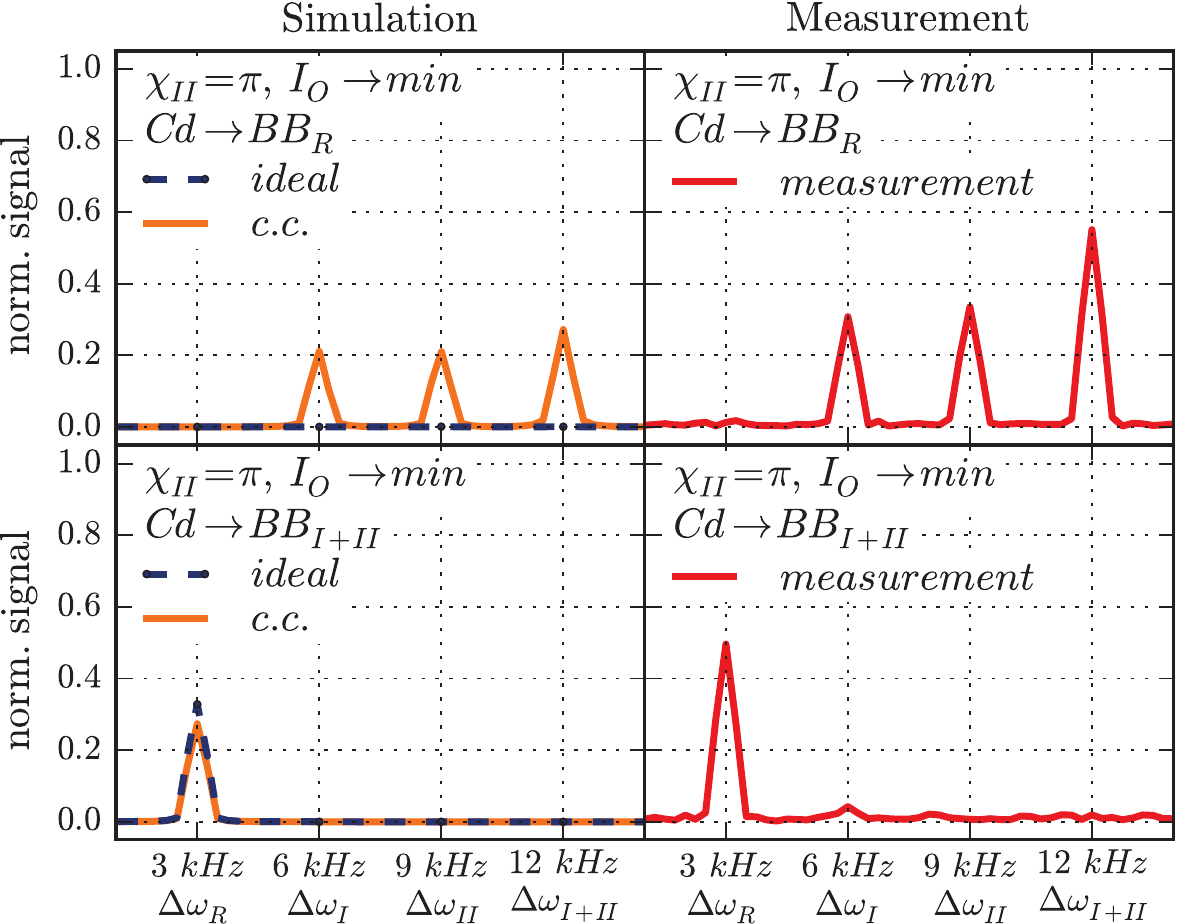}
 	 \caption{(color online). Power spectra for the phase setting $\chi_{II}=\pi$. In the upper row a beam-blocker in the position $BB_R$ (the reference beam is blocked); in the lower row a beam-blocker is put in the position $BB_{I+II}$ (the beam $I+II$ is blocked).}
\label{fig:Sectrum_CD}
\end{figure}

The time spectra $\Delta I$ are Fourier-analyzed in a standard manner using zero-padding and a Hanning windowing function. The power spectra are depicted in Figs.\ref{fig:NullPi} and \ref{fig:Sectrum_CD}.
The measurement results are shown on the right; simulations under ideal circumstances and
with corrected contrast (c.c.) are shown by dashed blue lines and solid orange lines,
respectively, on the left.
The ideal simulation is performed by calculating the intensities as given in Eq.\ref{eq:7}.
The c.c. simulation is carried out by adding contrast parameters $C_{i,j}$,
which denote actual (reduced) \modi{contrasts }{capability of the interference effect;
in practice, they are determined from the ratio between the amplitude of the interference fringes and the mean intensity} of interferograms.
The c.c. intensity is given by
\begin{equation}
I^\pm_{c.c.}=  \frac{1}{8} \sum\limits_{i,j} C_{i,j}\Psi_i^\ast\Psi_{j}
+ 2\sum\limits_{i,k} C_{i,k} \  \mathfrak{Re} \left(\Psi_i^\ast \Psi^\pm_k \right)
\label{eq:10}
\end{equation}
with $C_{i,i}=1$. Note that $i,j=\{I,II,R\}$ and $k=\{I,II,I+II,R\}$. Here, the first and the second terms represent stationary and the sinusoidally-oscillating intensities, respectively.
We set the contrast parameters for each pair of paths,
$C_{I,II}=C_{II,I}=0.55$, $C_{I,R}=C_{R,I}=0.60$, and $C_{II,R}=C_{R,II}=0.5$, as measured in the experiment.
\modi{}{Note that, the contrast parameters $C \ne 1$ infer \textit{imperfect} interference capacity of the interferometer.}
The intensity differences $\Delta I$ of both simulations are Fourier transformed in the same manner as the experimental data. All graphs are normalized to account for varying measurement times.

In the upper row of Fig.\ref{fig:NullPi}, power spectra of simulations and the measurement are depicted for the phase setting $\chi_{II}=0$. We find all peaks at the expected frequencies, $\Delta \omega_R=3$\,kHz, $\Delta \omega_I=6$\,kHz, $\Delta \omega_{II}=9$\,kHz, and $\Delta \omega_{I+II}=12$\, kHz, indicated by vertical grid lines. The ideal simulation, the c.c. simulation, and the measurement have the same peak heights for the respective frequencies. The peaks at frequencies $\Delta \omega_R$, $\Delta \omega_I$, and $\Delta \omega_{II}$ are the same height, while $\Delta \omega_{I+II}$ is twice the height. Since the WW-signal from SR$_{I+II}$ has to pass one beam splitter less than the signals from SR$_{I}$, SR$_{II}$, and SR$_{R}$ on the way to the detector, its amplitude is larger by a factor of $\sqrt{2}$. (see $a_i$ in Eq.\ref{eq:6}) Additionally, since SR$_{I+II}$ marks the superimposed wave functions $(\Psi'_I+\Psi'_{II})/\sqrt{2}$, the amplitude of the signal from SR$_{I+II}$ gains another factor of $\sqrt{2}$. The resulting amplitude of the signal of SR$_{I+II}$ has twice the amplitude of the signals from the other SRs. \modi{This causes, that}{Consequently,} the height of the peak at frequency $\Delta \omega_{I+II}$ is twice the height the other peaks in the power spectra.
This can also been seen in Eq.\ref{eq:8}.

The power spectrum with the phase setting $\chi_{II}=\pi$ is shown in the lower row of Fig.\ref{fig:NullPi}.
From Eq.\ref{eq:9}, one expects two peculiarities for the ideal simulation (dashed blue line). (i) The signal of SR$_{I+II}$ is zero and no peak is visible at frequency $\Delta \omega_{I+II}$, since SR$_{I+II}$ marks the wave function $(\psi_{I}+\E^{\i \pi}\psi_{II})/\sqrt{2}=0$.
(ii) The peaks at frequencies $\Delta \omega_R$, $\Delta \omega_I$ and $\Delta \omega_{II}$ drop to one third, since \textit{the amplitude} of $\Psi_{E_0}$ becomes $1/3$ (see also the description just after Eq.\ref{eq:9}). \modi{Not}{Note} that all these peculiarities can be validated by simply calculating \modi{}{the} cross-terms in Eq.\ref{eq:7}.

\modi{In our experiment, the beam in the path $I+II$ has non-zero intensity due to imperfect destructive-interference by $C_{I,II}<1$. The leakage of the front loop, i.e.,
$\Psi_{I+II} \ne 0$ leads to the non-zero WW-signal by $SR_{I+II}$; the peak at frequency $\Delta \omega_{I+II}$ emerges.}
{In ideal circumstances, the beam in the path $I+II$ is expected to have (absolutely) zero intensity due to destructive interference. Nevertheless, \textit{imperfect} destructive-interference in our experiment,
which is represented by $C_{I,II}<1$, results in non-zero intensity of this beam,
i.e., $\Psi_{I+II} \ne 0$; one finds a leakage of neutrons (in the main component) from the front loop to the beam path $I+II$. This leakage leads to the non-zero WW-signal by $SR_{I+II}$; the peak at frequency $\Delta \omega_{I+II}$ emerges as a consequence of the sum of cross terms
$\sum\limits_i C_{i,I+II} \  \mathfrak{Re} \left(\Psi_i^\ast \Psi^\pm_{I+II} \right)$ in Eq.\ref{eq:10}.}
In addition, the practical contrasts \modi{$C_{i,j} \ne 1$ }{$C_{i,j} < 1$}leads to changes of
peak heights at $\Delta \omega_i (i=R, I, II)$ in the power spectrum,
as seen in the lower row of Fig.\ref{fig:NullPi}.
Note that the intensities from the cross-terms between $\Psi^\pm_j$ and $\Psi^\ast_I$, $\Psi^\ast_R$ [$\Psi^\ast_{II}$] modulate sinusoidally in phase [out of phase due to $\chi_{II}=\pi$]
(see Eq.\ref{eq:10}).
That is, while \modi{beams in the same path exhibit full interference, i.e., $C_{i,i}=1$, and two}
{most of the cross terms in the last term of Eq.\ref{eq:10} contribute constructively, some} oscillating terms \modi{in Eq.\ref{eq:10} are $\pi$ phase-shifted due to $\chi_{II}=\pi$.}{there, i.e., from the cross terms between $\Psi_{II}$ and the others, do destructively.}
The peaks at the frequencies $\Delta \omega_R$, $\Delta \omega_I$ and $\Delta \omega_{II}$ are not equal in
height any more. This estimate is confirmed by the c.c. simulation, which agrees well with obtained results; slight deviation from the simulation is considered due to instability of the interferometer setup, i.e., phase and contrast fluctuation during the measurements.

In comparison, a beam-blocker, made of $1$\,mm thick Cadmium, is inserted in a position $BB_R$ or $BB_{I+II}$. Power spectra of simulations and the measurement are depicted in Fig.\ref{fig:Sectrum_CD}.
With the beam-blocker in the position $BB_R$ (upper row), the beam in path $R$ no longer contributes.
The ideal simulations show no peak at all; since $\Psi_{E_0}=0$, the mean intensity \modi{$I^{\pm}_0$}{$I^{\pm}_{E_0}$}
and all cross-terms $\Psi^{\ast}_{E_0}\Psi^{\pm}_i$ become zero (up to the first order).
Furthermore, the peak at frequency $\Delta \omega_R$ vanishes since all components in path $R$ are blocked.
In our measurement, due to \modi{leakage}{the leakage of the main component} from the front loop,  \modi{i.e., contrast of $C_{I,II}<1$}{resulting from $C_{I,II}<1$},
the peaks at frequencies $\Delta \omega_I$, $\Delta \omega_{II}$, and $\Delta \omega_{I+II}$ emerge
\modi{.}{ as a consequence of the sum of cross terms
$\sum\limits_i C_{i,j} \  \mathfrak{Re} \left(\Psi_i^\ast \Psi^\pm_{j} \right)$ with $(i,j) = (\{I,II\},\{I,II,R\})$ in Eq.\ref{eq:10}.
}
When the beam-blocker is put in the position $BB_{I+II}$ the WW-signals of SR$_I$, SR$_{II}$, and SR$_{I+II}$ are blocked and the respective peaks are invisible in the power spectrum (lower row). The height of the peak at frequency $\Delta \omega_{R}$ is unchanged
by considering reduced contrasts or not, since the power signal at $\Delta \omega_{R}$ arises from
the \modi{unchanged}{unaffected} cross-term $\Psi^{\ast}_{R}\Psi^{\pm}_{R}$.
The low count rate at this setting causes large statistical fluctuations in the measurement,
which is attributed to the increased height of the peak at frequency $\Delta \omega_{R}$.
The c.c simulation well reproduces the results of the measurement.

\section{IV. DISCUSSIONS}

The theoretical treatment, presented here, is obtained completely from the standard formalism of quantum mechanics. Predictions of the standard formalism agree well with the experimental results;
features emerging in the power spectrum, i.e., WW-information, are
indeed validated as the consequences of the interfering cross-terms between the main energy-unshifted component and the WW-signals. Note that, the WW-signal, being tiny in intensity, brings about
remarkably large outcomes. This is due to the fact that the heights of the peaks in the power spectrum
are proportional to the \textit{amplitude} of the WW-signal, which in turn, is given by the square-root of the intensity; the tiny signal still has relatively large amplitude \cite{SRT}.
The present experiment confirms the fact that the standard quantum mechanics does provide
an intuitive picture as well as a correct quantitative argument of the observed phenomena
in the three-beam interferometer.
Against the author's claim  of the letter \cite{22}, this is done in a far more appropriate manner than that by the use of  the two-state vector formulation; in particular, the latter formulates no quantitative justification.
Furthermore, our experiments demonstrated remarkable consequences of the destructive interference in practical circumstances, which is depicted in the simulation plotted in the lower row of Fig.\ref{fig:NullPi}.
First, the emergence of a WW-information, i.e., at the frequency of $\Delta \omega_{I+II}$, resulting from the leakage of the beam due to \modi{\textit{non-perfect}}{\textit{imperfect}} destructive interference. Next, we expect almost vanishing WW-information, i.e., at the frequency of $\Delta \omega_{II}$, by considering the practical contrasts.
This is due to the fact that one constructive and two destructive contribution (the latter is weakened by the reduced contrasts) in the terms
$\modi{2}{} \sum\limits_{i} C_{i,II} \  \mathfrak{Re} \left(\Psi_i^\ast \Psi^\pm_{II} \right)$ in Eq.\ref{eq:10},
are almost counterbalanced; along with the phase shifter positions $\chi_i$, the contrasts $C_{i,j}$
tune the extent of each contribution to the final WW-information.
Finally, the present experiment studies the (complete) absence of the wave function, corresponding to no propagating neutrons, by putting the beam block in the beam. In this case, sub-components, i.e., $\Psi_I$ and $\Psi_{II}$, as well as all WW-signals in that beam are blocked, which is confirmed again by the experiment.
%Detailed description together with a energy diagram as well as the obtained time spectra of the intensity modulating signals are given in supplemental Material.%

In our \modi{first}{} experiment, we study \modi{}{first} two phase settings, (i) $\chi_{II}=0$ and (ii) $\chi_{II}=\pi$,
both with $\chi_{I}=\chi_R=0$. In ideal circumstances, i.e, $C_{i,j}=1$, the wave function of the interfering beam $I+II$ is calculated as the consequence of \modi{}{full} constructive and destructive interference respectively, i.e.,
$\psi_{I+II} = \frac{1}{\sqrt{2}}(\psi_I+\psi_{II})$ for (i) and
$\psi_{I+II} = \frac{1}{\sqrt{2}}(\psi_I-\psi_{II})=0$ for (ii).
Note that, in the latter, while the intensity of the energy-unshifted component $\Psi_{I+II}$ is zero,
the WW-signals from the $SR_I$ and $SR_{II}$ remain, unaffected by the interference effect,
alive in the beam $I+II$;
while \modi{the beams in the same energy levels lead to}{the components in the original energy levels are affected by} (stationary) destructive interference effect, the \modi{beams}{components} in different \modi{every levels exhibit interference in time}{energy levels pass through the interferometer loop to cause the final sinusoidal intensity modulations in time} \cite{Badi} (see also the energy diagram depicted in Fig.\ref{fig:setup}(b)).
This is the reason why the WW-signal of the path I and II propagate through the beam path $I+II$,
although that of the path $I+II$, stemming from $\Psi_{I+II}$, is zero.
In contrast, a different situation occurs when the beam in the path $I+II$ is completely stopped by inserting the beam blocker in that beam, i.e., in the position $BB_{I+II}$. In this case, sub-components, $\Psi_I$ and $\Psi_{I+II}$ of the main beam together with both WW-signals from the $SR_I$ and $SR_{II}$ are blocked; there is no contribution, in the final O-beam, from these components any more.
Furthermore, in studying the interference effect, particularly in a (completely) destructive case, zero intensity appears; this situation is interpreted in a mistaken manner as non-continuous trajectories in \cite{22}. Appropriate consideration should be derived as the limit of (practically feasible) circumstances, i.e. $\Delta \chi \sim \pi$ and/or $C_{i,j} \sim 1$ (they are not exactly but nearly equal to $\pi$ and $1$). There propagate (smaller and smaller numbers but still some) quantum particles; they are actually there. \modi{In the present experiment, moderate situations appeared}{Moderate situations appear in the present experiment,} due to reduced contrasts $C \ne 1$ of the interferometer loops.

It is instructive to show the propagation of a WW-signal through the interferometer circuit, here.
When a WW-marking is done in one of the paths of an interferometer loop, the WW-signal is unaffected by the interference effect, as explained above, and only splitted into two outgoing beams at the \modi{}{eventual} beam splitter. Rather different circumstances emerge, when a WW-marking is done on the beam,
particularly before it \modi{entering}{enters} the interferometer circuit.
In this case, the WW-signals go through the interferometer loops:
\modi{it splits}{they split} into two beams in the interferometer and leave the interferometer
to \modi{appear}{emerge} in the two outgoing beams depending on the phase relation of the two interfering beams.
That is, by tuning the phase setting at the destructive interference position, i.e., the phase difference of $\pi$, the WW-signal does not \modi{appear}{emerge} in the interfering beam, say in the forward direction;
the WW-signal before the interferometer is only redireted through the interferometer loop.
The situation that \modi{the WW-signal is not found in the final detector does not necessarily mean that the
WW-signal does not exist at the WW marking}{no WW-signal is found in the final detector does not necessarily mean that no WW-signal has existed at the position of the WW marking};
these signal may be redirected to other beams.
This situation exactly happened in an experiment by Danan et al. (see Fig.3 in \cite{22});
the WW-signal at the mirror E is only redirected by the interferometer circuit and
does not arrive at the detector.

\section{V. CONCLUSION}

\modi{In summary, we }{We} have presented an interferometer experiment with massive particles to perform a kind of \modi{simultaneous}{multiple} which-way measurement with minimal-perturbations. Following the time evolution of the wave function of neutrons propagating through the interferometer, WW-signal with a time-dependent phase has been calculated. We have explicitly shown that the sinusoidally-oscillating intensities, from which WW-information is derived, are attributed to the cross terms between the main energy-unshifted component and the WW-signals. Experimental results \modi{agreed}{agree} well with \modi{the theoretical predictions}{the predictions to confirm the validity of our treatment in the framework of standard quantum mechanics}.
The present experiment witnesses the multifold presence of the neutron's wave function in the interferometer.
In addition, it gives clear justification of the utilized method, i.e., path extraction from the faint traces, and new insights of operation meaning of "the particle's path" at the quantum level.

\section{ACKNOWLEDGMENTS}

This work was supported by the Austrian Science Fund (FWF) Project  Nos. P24973-N20 \& P25795-N20. In addition we thank the ILL-Grenoble for its hospitality and continuous support.

\end{document}